\documentclass{article}

\usepackage{arxiv}

\usepackage[utf8]{inputenc} 
\usepackage[T1]{fontenc}    
\usepackage{url}            
\usepackage{booktabs}       
\usepackage{amsfonts}       
\usepackage{nicefrac}       
\usepackage{microtype}      
\usepackage[colorlinks = TRUE, linkcolor=blue,citecolor=blue,urlcolor=blue]{hyperref}
\usepackage{doi}

\usepackage{mathrsfs, amssymb, amsmath, amsthm, mathtools, graphicx, subfigure, psfrag, rotating, pdflscape, url, colortbl, multirow}
\usepackage[authoryear]{natbib}
\usepackage{cleveref}       

\title{The flexible Gumbel distribution: A new model for inference about the mode}

\date{\today}

\author{ \href{https://orcid.org/0000-0003-3265-6330}{\includegraphics[scale=0.06]{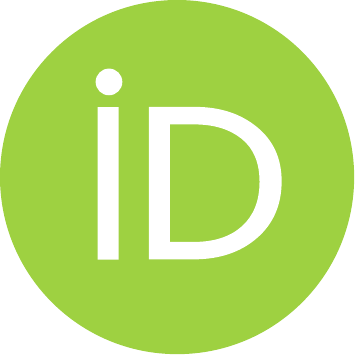}\hspace{1mm}Qingyang Liu} \\
	Department of Statistics\\
	University of South Carolina\\
	Columbia, SC 29201 \\
	\texttt{qingyang@email.sc.edu} \\
	\And
	\href{https://orcid.org/0000-0001-7077-0869}{\includegraphics[scale=0.06]{orcid.pdf}\hspace{1mm}Xianzheng Huang} \\
	Department of Statistics\\
	University of South Carolina\\
	Columbia, SC 29201 \\
	\texttt{huang@stat.sc.edu} \\
	\And
	\href{https://orcid.org/0000-0002-2777-5354}{\includegraphics[scale=0.06]{orcid.pdf}\hspace{1mm}Haiming Zhou} \\
	Daiichi Sankyo Inc.\\
	Basking Ridge, NJ 07920 \\
	\texttt{haiming2019@gmail.com} \\
}

\renewcommand{\shorttitle}{The flexible Gumbel distribution}

\hypersetup{
	pdftitle={The flexible Gumbel distribution: A new model for inference about the mode},
	pdfsubject={statistics},
	pdfauthor={Qingyang Liu, Xianzheng Huang and Haiming Zhou},
	pdfkeywords={extreme values, mixture distribution, modal regression, unimodal distribution},
}
\newcommand{\revise}{\color{black}}
\usepackage{fancyvrb}
\begin{document}
	\maketitle
	
\begin{abstract}
A new unimodal distribution family indexed by the mode and three other parameters is derived from a mixture of a Gumbel distribution for the maximum and a Gumbel distribution for the minimum. Properties of the proposed distribution are explored, including model identifiability and flexibility in capturing heavy-tailed data that exhibit different directions of skewness over a wide range. Both frequentist and Bayesian methods are developed to infer parameters in the new distribution. Simulation studies are conducted to demonstrate satisfactory performance of both methods. By fitting the proposed model to simulated data and data from an application in hydrology, it is shown that the proposed flexible distribution is especially suitable for data containing extreme values in either direction, with the mode being a location parameter of interest. {\revise Using the proposed unimodal distribution, one can easily formulate a regression model concerning the mode of a response given covariates. We apply this model to data from an application in criminology to reveal interesting data features that are obscured by outliers.} Computer programs for implementing all considered inference methods in the study are available at \href{https://github.com/rh8liuqy/flexible_Gumbel}{https://github.com/rh8liuqy/flexible\_Gumbel}.
\end{abstract}


\section{Introduction}
\label{s:intro}
The mean, median, and mode are {\color{black} the} three most commonly used measure of central tendency of data. When data contain outliers that cause heavy tails or are potentially skewed, the mode is a more sensible representation of the central location of data than the mean or median. The timely review on mode estimation and its application by \citet{Chacn2020} and references therein provide many examples in various fields of research where the mode serves as a more informative representative value of data. Most existing methods developed to draw inference for the mode are semi-/non-parametric in nature, starting from early works on direct estimation in the 1960s \citep{chernoff1964estimation, dalenius1965mode, venter1967estimation} to more recent works based on kernel density estimation \citep{chen2018modal} and quantile-based methods \citep{ota2019quantile, zhang2021bootstrap}. {\color{black} Two main factors contribute to the enduring preference for semi-/non-parametric methods for mode estimation, despite the typically less straightforward implementation and lower efficiency compared to parametric counterparts. First, parametric models often impose strict constraints on the relationship between the mode and other location parameters, which may not hold in certain applications.} Second, very few existing named distribution families that allow the inclusion of both symmetric and asymmetric distributions in the same family can be parameterized so that it is indexed by the mode as the location parameter along with other parameters, such as shape or scale parameters. In this study, we alleviate concerns raised by both reasons that discourage the use of parametric methods for mode estimation by formulating a flexible distribution indexed by the (unique) mode and parameters controlling the shape and scale. 

When it comes to modeling heavy-tailed data, the Gumbel distribution \citep{Gumbel1941} is arguably one of the most widely used models in many disciplines. Indeed, as a case of the generalized extreme value distribution \citep{jenkinson1955frequency}, the Gumbel distribution for the maximum (or minimum) is well-suited for modeling extremely large (or small) events that produce heavy-tailed data. For example, it is often used in hydrology to predict extreme rainfall and flood frequency  \citep{Loaiciga1999, Koutsoyiannis2004, Dawley2019}. In econometrics, the Gumbel distribution plays an important role in modeling extreme movements of stock prices and large changes in interest rates \citep{Bali2003, Pratiwi2019}. The Gumbel distribution is indexed by the mode and a scale parameter, and thus is convenient for mode estimation. However, the Gumbel distribution for the maximum (or minimum) is right-skewed (or left-skewed) with the skewness fixed at around {\color{black} $1.14$} (or {\color{black}$-1.14$}), and the kurtosis fixed at 5.4 across the entire distribution family. Thus it may be too rigid for scenarios where the direction and extremeness of outliers presented in data are initially unclear, or when the direction and level of skewness are unknown beforehand. Constructions of more flexible distributions that overcome these limitations have been proposed. In particular, \citet{cooray2010generalized} applied a logarithmic transformation on a random variable following the odd Weibull distribution to obtain the so-called generalized Gumbel distribution that includes the Gumbel distribution as a subfamily. But the mode of the generalized Gumbel distribution is not indexed by a location parameter, or an explicit function of other model parameters. \citet{Shin2015} considered mixture distributions with one of the components being the Gumbel distribution and the other component(s) being Gumbel of the same skewness direction or a different distribution, such as the gamma distribution. Besides the same drawback pointed out for the generalized Gumbel distribution, it is difficult to formulate a unimodal distribution following their construction of mixtures, and thus their proposed models are unsuitable when unimodality is a feature required to make inferring the mode meaningful, such as in a regression setting, as in modal regression \citep{yao2012local, Yao2013, chen2018modal}. 

With heavy-tailed data in mind and the mode as the location parameter of interest, we construct a new unimodal distribution that does not impose stringent constraints on how the mode relates to other central tendency measures, while allowing a range of kurtosis wide enough to capture heavy tails at either direction, as well as different degrees and directions of skewness. This new distribution, called the flexible Gumbel (FG) distribution, is presented in Section~\ref{s:FG}, where we study properties of the distribution and discuss identifiability of the model. We present a frequentist method and a Bayesian method for estimating parameters in the FG distribution in Section~\ref{s:inference}. Finite sample performance of these methods is inspected in simulation study in Section~\ref{s:simulation}, followed by an application of the FG distribution in hydrology in Section \ref{s:real}. Section~\ref{s:crime} demonstrates fitting a modal regression model based on the FG distribution to data from a criminology study. Section~\ref{s:discuss} highlights the contributions of our study and outlines future research directions.

\section{The flexible Gumbel distribution}
\label{s:FG}
The probability density function (pdf) of the Gumbel distribution for the maximum is given by 
\begin{align}
	f(x; \theta, \sigma) & =\frac{1}{\sigma} \exp\left\{-\frac{x-\theta}{\sigma}-\exp\left(-\frac{x-\theta}{\sigma}\right)\right\},
	\label{eq:RG}
\end{align}
where $\theta$ is the mode and $\sigma>0$ is a scale parameter. The pdf of the Gumbel distribution for the minimum with  mode $\theta$ and a scale parameter $\sigma$ is given by 
\begin{align}
	f(x; \theta, \sigma) & =\frac{1}{\sigma} \exp\left\{\frac{x-\theta}{\sigma}-\exp\left(\frac{x-\theta}{\sigma}\right)\right\}.
	\label{eq:LG}
\end{align}
We define a unimodal distribution for a random variable $Y$ via a mixture of the two Gumbel distributions specified by (\ref{eq:RG}) and (\ref{eq:LG}) that share the same mode $\theta$ while allowing different scale parameters, $\sigma_1$ and $\sigma_2$, in the two components. We call the resultant distribution the flexible Gumbel distribution, FG for short, with the pdf given by 
\begin{equation}
	\begin{aligned}
		f(y)=&\ w \times \frac{1}{\sigma_{1}} \exp \left\{-\frac{x-\theta}{\sigma_{1}}-\exp \left(-\frac{x-\theta}{\sigma_{1}}\right)\right\}+\\
		&\ (1-w) \times \frac{1}{\sigma_{2}} \exp \left\{\frac{x-\theta}{\sigma_{2}}-\exp \left(\frac{x-\theta}{\sigma_{2}}\right)\right\},
	\end{aligned}
	\label{eq:PDFFG}
\end{equation}
where \(w \in [0,1]\) is the mixing proportion parameter. Henceforth, we state that $Y\sim \mbox{FG}(\theta, \sigma_1, \sigma_2, w)$ if $Y$ follows the distribution specified by the pdf in (\ref{eq:PDFFG}). 

For each component distribution of FG, the mean and median are both some simple shift of the mode, with each shift solely determined by the scale parameter. Because the two components in (\ref{eq:PDFFG}) share a common mode $\theta$, the mode of $Y$ is also $\theta$, and thus the FG distribution is convenient to use when one aims to infer the mode as a central tendency measure, or to formulate parametric modal regression models \citep{bourguignon2020parametric, zhouhuang2020, zhou2022bayesian}. One can easily show that the mean of $Y$ is $E(Y) = w(\theta+\sigma_1\gamma) +(1-w)(\theta-\sigma_2\gamma) = \theta + \{w(\sigma_1+\sigma_2)-\sigma_2\}\gamma$, where $\gamma\approx 0.5772$ is the Euler-Mascheroni constant. Thus the discrepancy between the mode and the mean of FG depends on three other parameters that control the scale and shape of the distribution. The median of $Y$, denoted by $m$, is the solution to the following equation, 
$$w\exp\left\{-\exp\left(-\frac{m-\theta}{\sigma_1} \right)\right\}+(1-w)\left[1-\exp\left\{-\exp\left( \frac{m-\theta}{\sigma_2} \right)  \right\}  \right]=0.5.$$ Even though this equation cannot be solved for $m$ explicitly to reveal the median in closed form, it is clear that $m-\theta$ also depends on all three other parameters of FG. In conclusion, the relationships between the three central tendency measures of FG are more versatile than those under a Gumbel distribution for the maximum or a Gumbel distribution for the minimum. 

The variance of $Y$ is $V(Y) = \{w\sigma_1^2 + (1-w)\sigma_2^2\}\pi^2/6 + w(1-w)(\sigma_1+\sigma_2)^2\gamma^2$, which does not depend on the mode parameter $\theta$. Obviously, by setting $w=0$ or 1, $\mbox{FG}(\theta, \sigma_1, \sigma_2, w)$ reduces to one of the Gumbel components. Unlike a Gumbel distribution that only has one direction of skewness at a fixed level (of {\color{black} $\pm 1.14)$}, an FG distribution can be left-skewed, or right-skewed, or symmetric. More specifically, with the mode fixed at zero when studying the skewness and kurtosis of FG, one can show {\revise (as outlined in Appendix \ref{appendix:thired_central_moment})} that the third central moment of $Y$ is given by 
\begin{equation}
	w\bar w (\sigma_1+\sigma_2)^2 \gamma\left\{\gamma^2(\bar w-w)(\sigma_1+\sigma_2) +0.5\pi^2 (\sigma_1-\sigma_2)\right\}+2 \zeta(3)\left(w\sigma_1^3-\bar w\sigma_2^3\right),
	\label{eq:m3}
\end{equation}
where $\bar w=1-w$, and $\zeta(3)\approx 1.202$ is Ap\'ery's constant. Although the direction of skewness is not immediately clear from  (\ref{eq:m3}), one may consider a special case with $w=0.5$ where (\ref{eq:m3}) reduces to $(\sigma_1-\sigma_2)\{\gamma\pi^2(\sigma_1+\sigma_2)^2/8+\zeta(3)(\sigma_1^2+\sigma_1\sigma_2+\sigma_2^2)\}$. Now one can see that $\mbox{FG}(\theta, \sigma_1, \sigma_2, 0.5)$ is symmetric if and only if $\sigma_1=\sigma_2$, and it is left-skewed (or right-skewed) when $\sigma_1$ is less (or greater) than $\sigma_2$. The kurtosis of $Y$ can also be derived straightforwardly, with a more lengthy expression than (\ref{eq:m3}) that we omit here, which may not shed much light on its magnitude except that it varies as the scale parameters and the mixing proportion vary, instead of fixing at 5.4 as for a Gumbel distribution. An R Shiny app depicting the pdf of $\mbox{FG}(\theta, \sigma_1, \sigma_2, w)$ with user-specified parameter values is available at \url{https://qingyang.shinyapps.io/gumbel_mixture/}, created and maintained by the first author. Along with the density function curve, the Shiny app provides skewness and kurtosis of the depicted FG density. From there one can see that the skewness can be much lower than {\color{black} $-1.14$} or higher than {\color{black} $1.14$}, and the kurtosis can be much higher than 5.4, suggesting that inference based on FG can be more robust to outliers than when a Gumbel distribution is assumed for data at hand, without imposing stringent assumption on the skewness of the underlying distribution. 

The flexibility of a mixture distribution usually comes with concerns relating to identifiability \citep{teicher1961identifiability, Teicher1963, yakowitz1968identifiability}. In particular, there is the notorious issue of label switching when fitting a finite mixture model \citep{redner1984mixture}. Take the family of two-component normal mixture (NM) distributions as an example, defined by $\{\mbox{NM}(\mu_1, \sigma_1, \mu_2, \sigma_2, w): \, w\mathcal{N}(\mu_1, \sigma_1^2)+(1-w)\mathcal{N}(\mu_1, \sigma_2^2), \mbox{ for $\sigma_1, \sigma_2>0$ and $w\in [0, 1]$}\}$. When fitting a data set assuming a normal mixture distribution, one cannot distinguish between, for instance, $\mbox{NM}(1, 2, 3, 4, 0.2)$ and $\mbox{NM}(3, 4, 1, 2, 0.8)$, since the likelihood of the data is identical under these two mixture distributions. As another example, for data from a normal distribution, a two-component normal mixture with two identical normal components and an arbitrary mixing proportion $w\in [0, 1]$ leads to the same likelihood, and thus $w$ cannot be identified. \citet{Teicher1963} showed that imposing a lexicographical order for the normal components resolves the issue of non-identifiability, which also excludes mixtures with two identical components in the above normal mixture family. Unlike normal mixtures of which all components are in the same family of normal distributions, the FG distribution results from mixing two components from different families, i.e., a Gumbel distribution for the maximum and a Gumbel distribution for the minimum, with weight $w$ on the former component. By construction, FG does not have the label-switching issue. And, {\revise we show in Appendix~\ref{appendix:identifiability} by invoking Theorem 1 in \citet{Teicher1963} that} the so-constructed mixture distribution is always identifiable even when the true distribution is a (one-component) Gumbel distribution. 

\section{Statistical inference}
\label{s:inference}
\subsection{Frequentist inference method}
\label{s:freq}
Based on a random sample of size $n$ from the FG {\color{black} distribution}, $\mathbf{y}=\{y_i\}_{i=1}^n$, 
maximum likelihood estimators (MLE) of all model parameters in $\mathbf{\Omega}=(\theta, \sigma_1, \sigma_2, w)$ can be obtained via the expectation-maximization (EM) algorithm \citep{Dempster1977}. To apply the EM algorithm, we introduce a latent variable \(Z\) that follows Bernoulli(\(w\)) such that the joint likelihood of $(Y,Z)$ is
\begin{equation}
	f_{Y,Z}(y, z)=\{w f_1(y; \theta, \sigma_1)\}^z \{(1-w)f_2(y; \theta, \sigma_2)\}^{1-z},
	\label{eq:complete_density_function}
\end{equation}
where \(f_1(y; \theta, \sigma_1)\) is the pdf in (\ref{eq:RG}) evaluated at $y$ with the scale parameter  $\sigma=\sigma_1$, and \(f_2(y; \theta, \sigma_2)\) is the pdf in (\ref{eq:LG}) evaluated at $y$ with the scale parameter $\sigma=\sigma_2$. A random sample of size $n$ from Bernoulli($w$), $\mathbf{z}=\{z_i\}_{i=1}^n$, is viewed as missing data, and $\{(y_i, z_i)\}_{i=1}^n$ are viewed as the complete data in the EM algorithm. {\revise It can be shown \citep[][Section 2.6.3a]{Boos2013} that integrating $z$ out from \eqref{eq:complete_density_function} indeed gives the density of $Y$ in \eqref{eq:PDFFG}. The log-likelihood based on the density in \eqref{eq:PDFFG} is usually not well-behaved as an objective function to be maximized with respect to $\mathbf{\Omega}$. By considering the complete-data log-likelihood based on \eqref{eq:complete_density_function}, one can often maximize an objective function that is better-behaved as we demonstrate next. More specifically,} the complete-data log-likelihood is 
\begin{equation}
	\ell(\mathbf{\Omega};\mathbf{y},\mathbf{z}) = \sum_{i=1}^n \{z_i \log(w f_1(y_i; \theta, \sigma_1)) + (1-z_i)\log((1-w) f_2(y_i; \theta, \sigma_2))\}.\label{eq:complkh}
\end{equation}

Starting from an initial estimate of $\mathbf{\Omega}$ (at the zero-th iteration), denoted by $\mathbf{\Omega}^{(0)}$, one iterates two steps referred to as the E-step and the M-step until a convergence criterion is met. In the E-step at the \((t+1)\)-th iteration, one computes the conditional expectation of (\ref{eq:complkh}) given $\mathbf{y}$ while assuming the true parameter value to be $\mathbf{\Omega}^{(t)}=(\theta^{(t)}, \sigma_1^{(t)}, \sigma_2^{(t)}, w^{(t)})$, that is, $\operatorname{E}_{\boldsymbol{\Omega}^{(t)}} \{\ell(\mathbf{\Omega}; \mathbf{y}, \mathbf{z})|\mathbf{y}\}$. This conditional expectation can be shown to be 
\begin{equation}
	Q\left(\boldsymbol{\Omega} \left\vert \mathbf{\Omega}^{(t)}\right.\right)= \sum_{i=1}^{n}\left\{T _ { i } ^ { ( t ) } \operatorname { l o g } (w f_{1}\left(y_{i} ; \theta, \sigma_{1}\right))+\left(1-T_{i}^{(t)}\right) \log ((1-w) f_{2}\left(y_{i} ; \theta, \sigma_{2}\right))\right\},
	\label{eq:Qfunction}
\end{equation}
where 
\begin{equation}
	T_i^{(t)} =  \operatorname{E}_{{\boldsymbol{\Omega}}^{(t)}}(Z|Y=y_i)=
	\frac{w^{(t)}f_1(y_i; \theta^{(t)}, \sigma_1^{(t)})}{w^{(t)}f_1(y_i; \theta^{(t)}, \sigma_1^{(t)})+(1-w^{(t)})f_2(y_i; \theta^{(t)}, \sigma_2^{(t)})}. \label{eq:Ti}
\end{equation}
In the M-step at the \((t+1)\)-th iteration, one maximizes \(Q(\boldsymbol{\Omega} | \boldsymbol{\Omega}^{(t)})\) with respect to $\mathbf{\Omega}$ to obtain an updated estimate $\mathbf{\Omega}^{(t+1)}=(\theta^{(t+1)}, \sigma_1^{(t+1)}, \sigma_2^{(t+1)}, w^{(t+1)})$, in which $w^{(t+1)} =\sum_{i=1}^n T_i^{(t)}/n$, and the other three updated estimates in $\mathbf{\Omega}^{(t+1)}$ are obtained numerically.   

{\revise The EM algorithm avoids directly maximizing the log-likelihood based on  \eqref{eq:PDFFG} by (iteratively) maximizing the better-behaved $Q(\mathbf{\Omega}|
	\mathbf{\Omega}^{(t)})$ in \eqref{eq:Qfunction}. To further improve the numerical efficiency, we exploit} the expectation-conditional maximization (ECM) algorithm \citep{MENG1993}, which replaces the M-step with a sequence of 
simpler conditional maximizations referred to as the CM-step. {\revise Essentially, within each $M$-step, we update $w$ via 
	$w^{(t+1)} =\sum_{i=1}^n T_i^{(t)}/n$, then we update $\theta$ using $w^{(t+1)}$ along with $(\sigma_1^{(t)}, \sigma_2^{(t)})$, followed by updating $\sigma_1$ using $w^{(t+1)}$, the recently updated $\theta$, and $\sigma_2^{(t)}$; lastly, we update $\sigma_2$ using $w^{(t+1)}$ and the recently updated $\theta$ and $\sigma_1$.} 
There is no closed-form updating formula for {\revise the latter three updates, but each of them} can now be easily updated by most well-accepted one-dimensional optimization algorithms, such as the  Newton-Raphson method. To ensure convergence at the global maximum, as recommended by \citet{Wu1983}, one should implement the ECM algorithm several rounds with different starting values  $\mathbf{\Omega}^{(0)}$. 

After obtaining the MLE of $\mathbf{\Omega}$, denoted by $\hat {\mathbf{\Omega}}$, we propose to use the sandwich variance estimator \citep[][Chapter 7]{Boos2013} to estimate the variance-covariance matrix of $\hat{ \mathbf{\Omega}}$. One may also estimate the variance-covariance of $\hat{\mathbf{\Omega}}$ based on the observed information matrix as described in \citet{Louis1982} and \citet{Oakes1999}. The benefit of using the sandwich variance estimator is its robustness to model misspecification. Finally, the EM and ECM algorithms bear a strong resemblance to
data augmentation \citep{wei1990monte} in the Bayesian framework, which we turn to next for inferring $\mathbf{\Omega}$. 

\subsection{Bayesian inference method}\label{sec:bayes_infer}
\label{s:bayes}
In the Bayesian framework, we formulate hierarchical models starting with the FG distribution, 
\[Y | \theta, \sigma_1,\sigma_2,w \sim \text{FG}(\theta, \sigma_1,\sigma_2, w),\]
followed by independent weakly informative or non-informative priors for elements in $\mathbf{\Omega}$,
\begin{align*}
	\theta & \sim \mathcal{N}(0,10^4),\\
	\sigma_j & \sim \text{inv-Gamma}(1,1), \text{ for } j = 1, 2,\\
	w & \sim \text{\color{black} Uniform}(0,1),
\end{align*}
where inv-Gamma refers to the inverse Gamma distribution. We choose the above prior for the scale parameters by following the prior selection for variance parameters suggested in \citet{Gelman2006}.

We employ the Metropolis-within-Gibbs sampler \citep{Muller1991, Muller1993} to obtain an estimate of $\mathbf{\Omega}$ from the posterior distribution of $\mathbf{\Omega}$ given observed data $\mathbf{y}$. Similar to the EM/ECM algorithm in Section~\ref{s:freq}, the latent variable \(Z\) is also introduced as a device to carry out data augmentation. The iterative algorithm presented next is based on the following two conditional distributions that can be easily proved, 
\begin{align*}
	z_i|\theta,\sigma_1,\sigma_2,w,\mathbf{z}_{-i},\mathbf{y} & \sim \text{Bernoulli}\left(\frac{w f_1(y_i;\theta,\sigma_1)}{w f_1(y_i;\theta,\sigma_1) + (1-w)f_2(y_i;\theta,\sigma_2)}\right),\\ w|\theta,\sigma_1,\sigma_2,\mathbf{z},\mathbf{y} & \sim \text{Beta}\left(1+\sum_{i=1}^n z_i, \, n+1-\sum_{i=1}^n z_i \right),
\end{align*}
where $\mathbf{z}_{-i}$ results from dropping $z_i$ from $\mathbf{z}$, and the first result above is also from which (\ref{eq:Ti}) is deduced.

The Metropolis-within-Gibbs sampler at the \((t+1)\)-th iteration involves four steps outlined below.
\begin{itemize}
	\item Step 1:  For \(i = 1,\dots,n\), draw $z^{(t+1)}_i$ from $\mbox{Bernoulli}(T_i^{(t)})$, where $T_i^{(t)}$ is given in (\ref{eq:Ti}).
	\item Step 2: Draw $w^{(t+1)}$ from  $\mbox{Beta}\left(1+\sum_{i=1}^{n} z^{(t+1)}_{i}, \, n+1-\sum_{i=1}^{n} z^{(t+1)}_{i}\right).$
	\item Step 3: Draw $\tilde\theta$ from  $\mathcal{N}(\theta^{(t)}, \tau_0)$, and update $\theta^{(t)}$ to $\theta^{(t+1)}$ according to the following decision rule, 
	\begin{equation*}
		\theta^{(t+1)} = 
		\begin{cases}
			\tilde\theta, \quad \text{with probability } \displaystyle{q=\min\left\{\frac{p(\tilde\theta|w^{(t+1)},\sigma^{(t)}_1, \sigma^{(t)}_2, \mathbf{y})}{p(\theta^{(t)}|w^{(t+1)},\sigma^{(t)}_1, \sigma^{(t)}_2, \mathbf{y})}, \,1\right\}},\\
			\theta^{(t)}, \quad \text{with probability $1-q$}.
		\end{cases}
	\end{equation*}
	\item Step 4: For $j=1, 2$, draw $\tilde\sigma_j$ from $ \mathcal{N}(\sigma_j^{(t)},\tau_j)$, and update $\sigma_j^{(t)}$ to $\sigma_j^{(t+1)}$ according to the following decision rule, for $k\ne j$,
	\begin{equation*}
		\sigma_j^{(t+1)} = 
		\begin{cases}
			\tilde\sigma_j, \quad \text{with probability } \displaystyle{q=\min\left\{\frac{p(\tilde\sigma_j|\theta^{(t+1)},\sigma^{(t)}_k,w^{(t+1)}, \mathbf{y})}{p(\sigma_j^{(t)}|\theta^{(t+1)},\sigma^{(t)}_k,w^{(t+1)}, \mathbf{y})}, \,1\right\}},\\
			\sigma_{j}^{(t)}, \quad \text{with probability $1-q$}.
		\end{cases}
	\end{equation*}
	
\end{itemize}
In Steps 3 and 4, \(p(\cdot | \cdot)\) refers to a conditional pdf generically, $\tau_0$, $\tau_1$, and $\tau_2$ are three positive tuning parameters whose values should be chosen so that the acceptance rate at each step is around \(23\%\) \citep{Gelman1997}. To draw samples from the joint posterior distribution, there are numerous ways to design the Markov chain Monte Carlo (MCMC) sampler. Instead of the Metropolis-within-Gibbs sampler we adopt here, one may use other existing MCMC software, such as \textsc{Stan} \citep{Rstan}, \textsc{JAGS} \citep{plummer2003jags}, and \textsc{BUGS} \citep{spiegelhalter1996bugs,lunn2009bugs}, {\color{black} the former two of which are demonstrated in the \href{https://github.com/rh8liuqy/flexible\_Gumbel}{GitHub repository (\texttt{https://github.com/rh8liuqy/flexible\_Gumbel})}}. After obtaining enough high-quality samples from the joint posterior distribution \(p(\theta,\sigma_1,\sigma_2,w | \mathbf{y})\), Bayesian inference is straightforward, including point estimation, interval estimation, and uncertainty assessment. 

\section{Simulation study}
\label{s:simulation}
Large-sample properties of MLEs and likelihood-based Bayesian inference under a correct model for data have been well studied. To assess finite-sample performance of the frequentist method and Bayesian method proposed in Section~\ref{s:inference}, we carried out simulation study with two specific aims: first, to compare inference results from the two methods; second, to compare goodness of fit for data from distributions outside of the FG family when one assumes an FG distribution and when one assumes a two-component normal mixture distribution for the data. 

In the first experiment, denoted as (E1) hereafter, {\revise we considered two FG distributions as true data generating mechanisms, $\text{FG}(\theta=1, \sigma_1=1, \sigma_2=1, w=0.4)$ and $\text{FG}(\theta=0, \sigma_1=1, \sigma_2=5, w=0.5)$. This design creates two FG distributions with the second one more skewed and variable than the first. Based on a random sample of size $n=50$ from the first FG distribution, we estimated $\mathbf{\Omega}$ by applying the ECM algorithm and the Metropolis-within-Gibbs algorithm. Similarly, based on a random sample of size $n \in \{100, 200\}$, we implemented the two algorithms to estimate $\mathbf{\Omega}$.} The former algorithm produced the MLE of $\mathbf{\Omega}$, and we used the median of the posterior distribution of $\mathbf{\Omega}$ at convergence of the latter algorithm as another point estimate of $\mathbf{\Omega}$. Table \ref{tab:E1tab} presents summary statistics of these estimates of $\mathbf{\Omega}$ and estimates of the corresponding standard deviation across 1000 Monte Carlo replicates {\revise under each simulation setting specified by the design of an FG distribution and the level of $n$}. 

\begin{table}
	\small
	\caption{\label{tab:E1tab}Frequentist and Bayesian inference results in experiment (E1) across 1000 Monte Carlo replicates. Here, point.est stands for the average of 1000 point estimates for each parameter from each method, $\widehat{\text{s.d.}}$  stands for the average of the corresponding 1000 estimated standard deviations, and s.d. refers to the empirical standard deviation of the 1000 point estimates from each method. Numbers in parentheses are $100\times$  Monte Carlo standard errors associated with the averages {\color{black} of 1000 estimates of the standard deviation associated with a point estimator}.}
	\centering
	\begin{tabular}[t]{llrlrrlr}
		\toprule
		\multicolumn{1}{c}{ } & \multicolumn{1}{c}{ } & \multicolumn{3}{c}{Frequentist} & \multicolumn{3}{c}{Bayesian} \\
		\cmidrule(l{3pt}r{3pt}){3-5} \cmidrule(l{3pt}r{3pt}){6-8}
		\multicolumn{1}{c}{sample size} & \multicolumn{1}{c}{parameter} & \multicolumn{1}{c}{point.est} & \multicolumn{1}{c}{$\widehat{\text{s.d.}}$} & \multicolumn{1}{c}{s.d.} & \multicolumn{1}{c}{point.est} & \multicolumn{1}{c}{$\widehat{\text{s.d.}}$} & \multicolumn{1}{c}{s.d.}\\
		\midrule
		& $\theta$ & 0.990 & 0.197 (0.40) & 0.209 & 0.965 & 0.250 (0.18) & 0.224\\
		\cmidrule{2-8}
		& $\sigma_1$ & 1.106 & 0.272 (0.77) & 0.419 & 1.045 & 0.638 (2.30) & 0.296\\
		\cmidrule{2-8}
		& $\sigma_2$ & 1.047 & 0.192 (0.49) & 0.216 & 1.082 & 0.459 (2.12) & 0.465\\
		\cmidrule{2-8}
		\multirow{-4}{*}{\raggedright\arraybackslash $n=50$} & $w$ & 0.411 & 0.190 (0.46) & 0.207 & 0.435 & 0.207 (0.12) & 0.187\\
		\cmidrule{1-8}
		& $\theta$ & 0.002 & 0.198 (0.20) & 0.201 & 0.013 & 0.205 (0.15) & 0.203\\
		\cmidrule{2-8}
		& $\sigma_1$ & 0.979 & 0.204 (0.41) & 0.216 & 1.014 & 0.224 (0.27) & 0.214\\
		\cmidrule{2-8}
		& $\sigma_2$ & 4.932 & 0.590 (0.56) & 0.613 & 4.813 & 0.666 (0.44) & 0.615\\
		\cmidrule{2-8}
		\multirow{-4}{*}{\raggedright\arraybackslash $n=100$} & $w$ & 0.495 & 0.091 (0.09) & 0.090 & 0.484 & 0.090 (0.04) & 0.088\\
		\cmidrule{1-8}
		& $\theta$ & 0.008 & 0.136 (0.08) & 0.129 & 0.011 & 0.137 (0.07) & 0.130\\
		\cmidrule{2-8}
		& $\sigma_1$ & 0.999 & 0.143 (0.21) & 0.144 & 1.013 & 0.144 (0.10) & 0.141\\
		\cmidrule{2-8}
		& $\sigma_2$ & 4.993 & 0.435 (0.32) & 0.431 & 4.940 & 0.457 (0.20) & 0.434\\
		\cmidrule{2-8}
		\multirow{-4}{*}{\raggedright\arraybackslash $n=200$} & $w$ & 0.500 & 0.064 (0.04) & 0.063 & 0.495 & 0.063 (0.02) & 0.062\\
		\bottomrule
	\end{tabular}
\end{table}

According to Table~\ref{tab:E1tab}, all parameter estimates in $\mathbf{\Omega}$ are reasonably close to the true values. {\revise When the sample size is as small as 50, estimates for $\mathbf{\Omega}$ resulting from the frequentist method are still similar to those from the Bayesian inference method, although estimates for the standard deviations of these point estimators can be fairly different. We do not find such discrepancy surprising because, for the frequentist method where we use the sandwich variance estimator to infer the uncertainty of an MLE for $\mathbf{\Omega}$, the asymptotic properties associated with MLEs that support the use of a sandwich variance estimator may not take effect yet at the current sample size; and, for the Bayesian method, the quantification of standard deviation can be sensitive to the choice of priors when $n$ is small. These are confirmed by the diminishing discrepancy between the two sets of standard deviation estimates when $n=100$, 200.} A closer inspection of the reported empirical mean of estimates for $\mathbf{\Omega}$ along with their empirical standard error suggests that, when $n=100$, the Bayesian method may slightly underestimate $\sigma_2$, the larger of the two scale parameters of FG. We believe that this is due to the inverse gamma prior imposed on the scale parameters that is sharply peaked near zero, and thus the posterior median of the larger scale parameter tends to be pulled downwards when the sample size is not large. As the sample size increases to $200$, this trend of underestimation appears to diminish. The empirical means of the standard deviation estimates from both methods are close to the corresponding empirical standard deviations, which indicate that the variability of a point estimator is accurately estimated {\revise when $n$ is not small}, whether it is based on the sandwich variance estimator in the frequentist framework, or based on the posterior sampling in the Bayesian framework. In summary, the methods proposed in Section~\ref{s:inference} under both frameworks provide reliable inference for $\mathbf{\Omega}$ along with accurate uncertainty assessment of the point estimators when data arise from an FG distribution.

{\color{black} Among all existing mixture distributions, normal mixtures probably have the longest history and are most referenced in the literature.} In another experiment, we compared the model fitting of normal mixture with that of FG when data arise from three heavy-tailed distributions: (E2) Laplace with the location parameter equal to zero and the scale parameter equal to 2; (E3) a mixture of two Gumbel distributions for the maximum, with a common mode at zero, scale parameters in the two components equal to 2 and 6, respectively, and the mixing proportion equal to 0.5; (E4) a {\color{black}Student-}$t$ distribution with degrees of freedom equal to 5. From each of the three distributions in (E2)--(E4), we generated a random sample of size $n=200$, following which we fit a two-component normal mixture model via the EM algorithm implemented using the R package  \texttt{mixtools}, and also fit an FG model via the two algorithms described in Section~\ref{s:inference}. This model fitting exercise was repeated for 1000 Monte Carlo replicates under each of (E2)--(E4). 

\begin{figure}
	\begin{center}
		\includegraphics[width=0.8\linewidth]{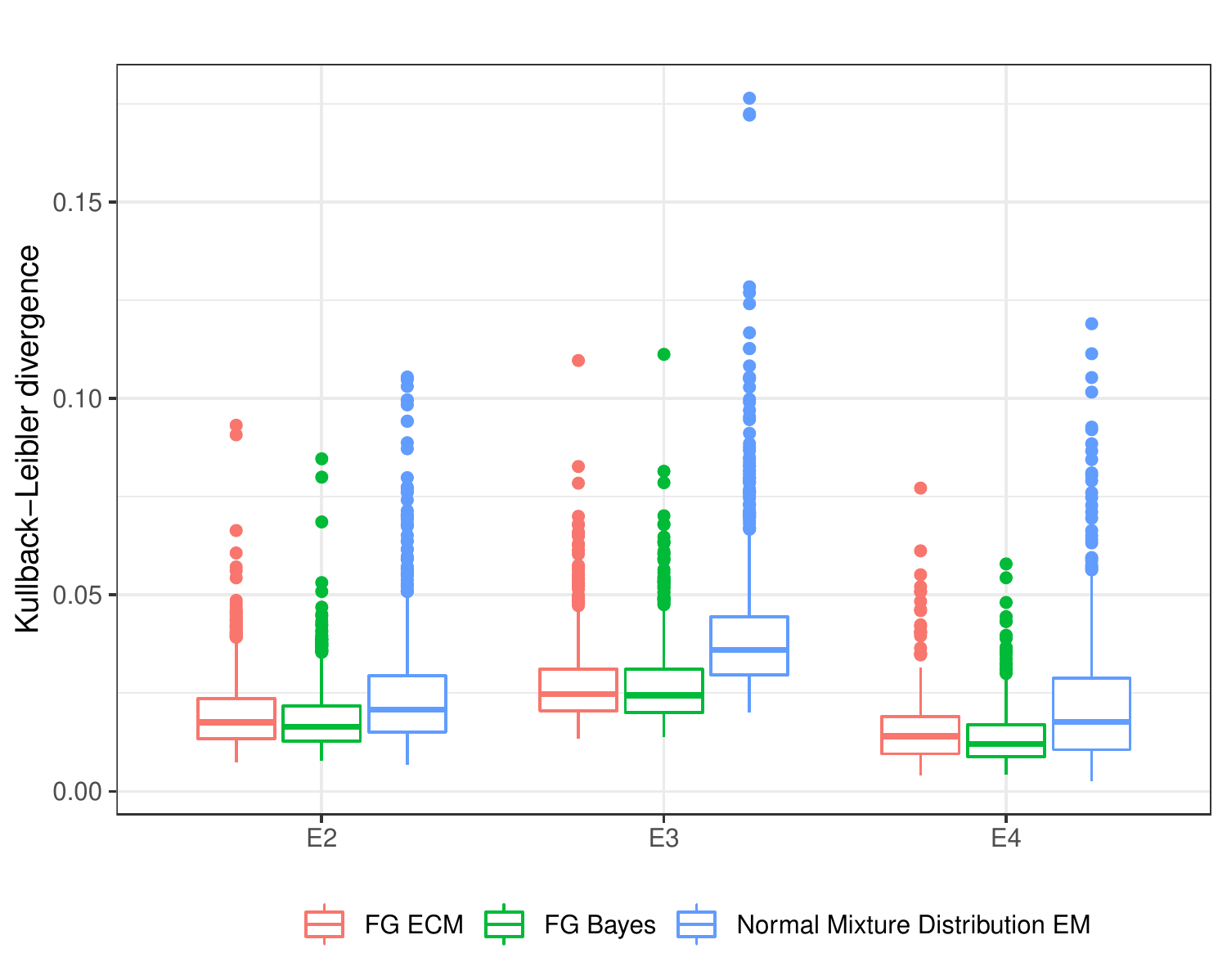}
	\end{center}
	\caption{\label{fig:KLdistance}Boxplots of the empirical Kullback-Leibler divergence from an estimated density to the true density under each of the true-model settings in  (E2)--(E4). Under each setting, the three considered model fitting strategies are, from left to right in the figure, (i) using the ECM algorithm to fit an FG distribution (FG ECM), (ii) using the Bayesian method to fit an FG distribution (FG Bayes), and (iii) using the EM algorithm to fit a normal mixture distribution (Normal Mixture Distribution EM).}
\end{figure}

We used an empirical version of the Kullback-Leibler divergence as the metric to assess the quality of modeling fitting. We denote the true density function as $p(\cdot)$, and let $\hat p(\cdot)$ be a generic estimated density resulting from one of the three considered model fitting strategies. Under each setting in (E2)--(E4), a random sample of size 50000, $(x_1, \ldots, x_{50000})$, was generated from the true distribution, and an empirical version of the Kullback-Leibler divergence from $\hat p(\cdot)$ to $p(\cdot)$ is given by \(D_{\hbox {\tiny\mbox{KL}}} = (1/50000)\sum_{i=1}^{50000} \log(p(x_i)/\hat{p}(x_i))\). Figure \ref{fig:KLdistance} shows the boxplots of $D_{\hbox {\tiny \mbox{KL}}}$ across 1000 Monte Carlo replicates corresponding to each model fitting scheme under (E2)--(E4). 

Judging from Figure~\ref{fig:KLdistance}, the FG distribution clearly {\revise outperforms} the normal mixture when fitting data from any of the three heavy-tailed distributions in (E2)--(E4), and results from the frequentist method are comparable with those from the Bayesian method for fitting an FG model. When implementing the ECM algorithm for fitting the FG model and the EM algorithm for fitting the normal mixture, we set a maximum number of iterations at 1000. Our ECM algorithm always converged in the simulation, i.e., converged to a stationary point within 1000 iterations. However, the EM algorithm for fitting a normal mixture often had trouble achieving that, with more difficulty when data come from a heavier-tailed distribution. More specifically, under (E4), which has the highest kurtosis (equal to 9) among the three settings, the EM algorithm failed to converge in 59.9\% of all Monte Carlo replicates; under (E2), which has the second highest kurtosis (equal to 6), it failed to converge in 6.7\% of the replicates. Results associated with the normal mixture from these failing replicates were not included when producing the boxplots in Figure~\ref{fig:KLdistance}. In conclusion, the FG distribution is more suitable for symmetric or asymmetric heavy-tailed data than the normal mixture distribution. 

\section{An application in hydrology}
\label{s:real}
Daily maximum water elevation changes of a water body, such as ocean, lake, and wetland, are of interest in hydrologic research. These changes may be close to zero on most days but can be extremely large or small under extreme weather. From {\color{black} the} National Water Information System (\url{https://waterdata.usgs.gov/}), we downloaded water elevation data for Lake Murray near Columbia, South Carolina, United States, recorded from September 18, 2020 to September 18, 2021. The water elevation change of a given day was calculated by contrasting the maximum elevation and the minimum elevation on that day, returning a positive (negative) value if the maximum record of the day comes after (before) the minimum record on the same day. We fit the FG distribution to the resultant data with $n=366$ records using the frequentist method and the Bayesian method, with results presented in Table~\ref{tab:lake}. The two inference methods produced very similar estimates for most parameters, although small differences were observed. For example, one would estimate the mode of daily maximum water elevation change to be $-0.795$ feet based on the frequentist method, but estimate it to be $-0.485$ feet using the Bayesian method. The discrepancy between these two mode estimates is minimal considering that the daily maximum water elevation changes range from $-38$ feet to 49.4 feet within this one year. Taking into account the uncertainty in these point estimates, we do not interpret any of these differences as statistically significant because a parameter estimate from one method always falls in the interval estimate for the same parameter from the other method according to Table~\ref{tab:lake}. Using parameter estimates in Table~\ref{tab:lake} in the aforementioned R Shiny app, we obtained an estimated skewness of $-0.102$ and an estimated kurtosis of 6.384 based on the frequentist inference results, whereas the Bayes inference yielded an estimated skewness of 0.058 and an estimated kurtosis of 6.074. Combining these two sets of results, we concluded that the underlying distribution of daily maximum water elevation change may be nearly {\color{black} symmetrical}, with outliers on both tails that cause tails heavier than that of a Gumbel distribution. 
\begin{table}
	\small
	\caption{\label{tab:lake}Frequentist and Bayesian inferences about daily maximum water elevation changes of Lake Murray, South Carolina, United States. Besides parameter estimates (under point.est) and the estimated standard deviations of these parameter estimates (under $\widehat{\mbox{s.d.}}$), 95\% confidence intervals of the parameters from the frequentist method, and 95\% credible intervals from the Bayesian method are also provided (under lower 95 and upper 95).}
	\centering
	\begin{tabular}[t]{lrrrrrrrr}
		\toprule
		\multicolumn{1}{c}{ } & \multicolumn{4}{c}{Frequentist} & \multicolumn{4}{c}{Bayesian} \\
		\cmidrule(l{3pt}r{3pt}){2-5} \cmidrule(l{3pt}r{3pt}){6-9}
		parameter & point.est & $\widehat{\text{s.d.}}$ & lower 95 & upper 95 & point.est & $\widehat{\text{s.d.}}$ & lower 95 & upper 95\\
		\midrule
		$\theta$ & -0.795 & 0.796 & -2.355 & 0.764 & -0.485 & 0.695 & -1.670 & 0.979\\
		$\sigma_1$ & 5.186 & 0.541 & 4.124 & 6.247 & 5.400 & 0.655 & 4.520 & 6.910\\
		$\sigma_2$ & 6.237 & 1.735 & 2.836 & 9.638 & 5.733 & 1.036 & 4.390 & 8.030\\
		$w$ & 0.698 & 0.169 & 0.367 & 1.029 & 0.629 & 0.141 & 0.327 & 0.846\\
		\bottomrule
	\end{tabular}
\end{table}

\begin{figure}
	\begin{center}
		\centering 
		\includegraphics[width=1.0\linewidth]{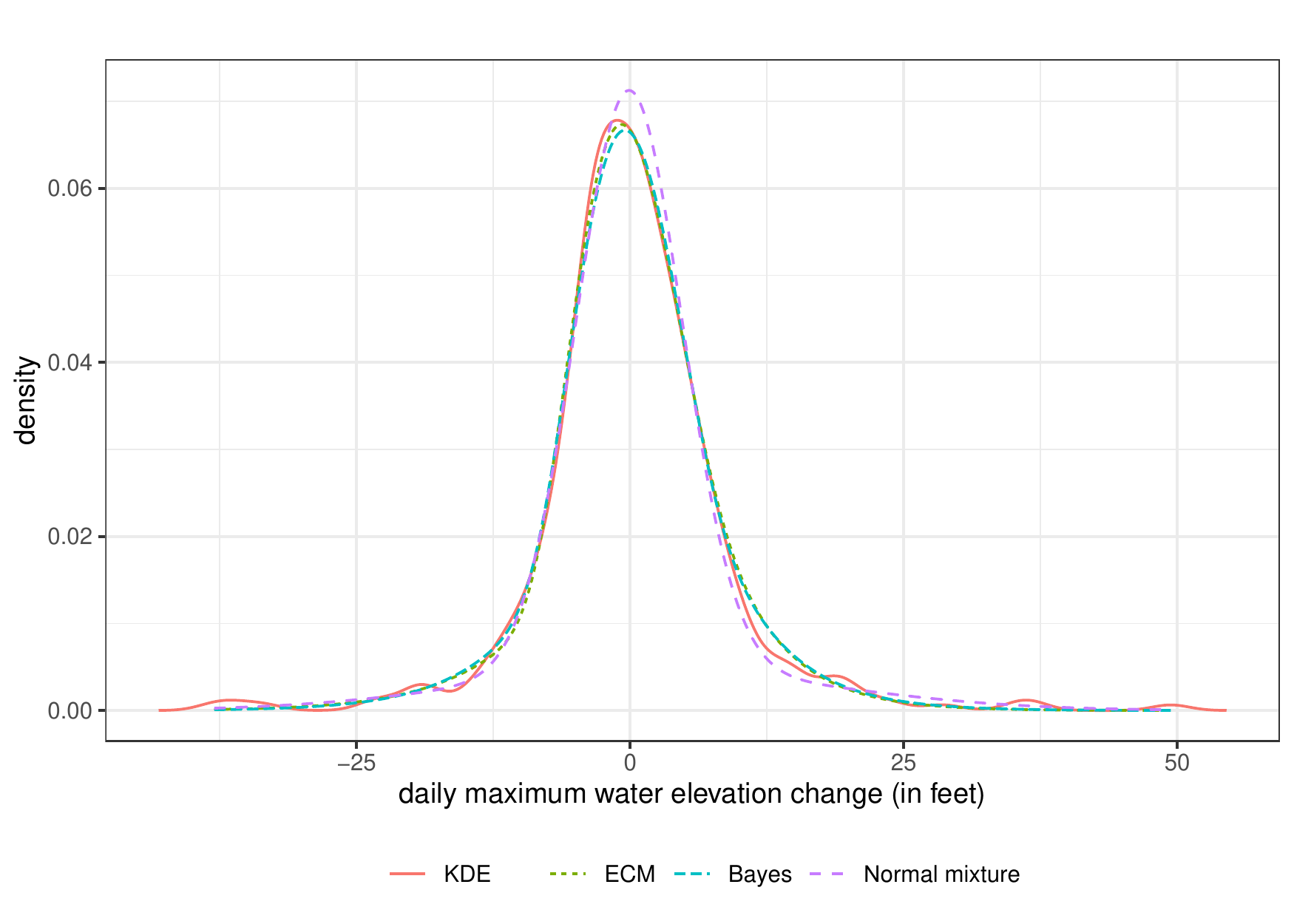} 
	\end{center}
	\caption{Four density estimates based on daily maximum water elevation changes in Lake Murray,  including the kernel density estimate (solid line), the estimated FG density from the ECM algorithm (dotted line), the estimated FG density from the Bayesian method (dashed line), and the estimated normal mixture density (dash-dotted line).}\label{fig:densityplot}
\end{figure}

Figure \ref{fig:densityplot} presents the estimated density functions from these two methods, in contrast with the estimated density curve resulting from fitting the data to a two-component normal mixture, and a kernel density estimate using a Gaussian kernel with the bandwidth selected according to the method proposed by \citet{oro28316}. 
The last estimate is fully nonparametric and served as a benchmark against which the other three density estimates were assessed graphically. The kernel density estimate is more flexible at describing varying tail behaviors, but such flexibility comes at the cost of statistical efficiency and interpretability. With the wiggly tails evident in Figure~\ref{fig:densityplot} for this estimate, we suspected a certain level of overfitting of the kernel density estimate. This often happens to kernel-based estimation of a function around a region where data are scarce, with a bandwidth not large enough for the region. Between the two FG density estimates, the difference is almost negligible. They both track the kernel density estimate closely over a wide range of the support around the mode. The mode of the estimated normal mixture density is close to the other three mode estimates, but the tails are much lighter than those of the other three estimated densities. 

Besides comparing the three parametric density estimates pictorially, we also used the Monte-Carlo-based one-sample Kolmogorov–Smirnov test to assess the goodness of fit. The $p$-values from this test are 0.223, 0.312, and 0.106 for the frequentist FG density estimate, the Bayesian FG density estimate, and the estimated normal mixture density, respectively. Although none of the $p$-values are low enough to indicate a lack of fit (at significance level 0.05 for example), the $p$-value associated with the normal mixture is much lower than those for FG. {\revise Hence, between the two null hypotheses, with one assuming an FG distribution and the other claiming a normal mixture for this data set, we find even weaker evidence to reject the former than data evidence against the latter.} It is also worth noting that the Kolmogorov–Smirnov test is known to have low power to detect deviations from a posited distribution that occur in the tails \citep{mason1983modified}. This may explain the above-0.05 $p$-value for the normal mixture fit of the data even though the tail of this posited distribution may be too thin for the current data. {\revise Finally, as suggested by a referee, we computed the Akaike information criterion (AIC) and the Bayesian information criterion (BIC) after fitting the FG distribution and the normal mixture distribution to the data. When assuming an FG distribution, we obtained an  AIC/BIC of 2506.028/2521.638 from the frequentist method, and 2506.299/2521.909 from the Bayesian method. When assuming a mixture normal, we found the values of AIC and BIC to be 2499.821 and 2519.334, respectively. Even though the fitted normal mixture distribution produces a lower AIC/BIC than the fitted FG distribution, we argue that these metrics focus more on the {\it overall} goodness of fit, and can be more forgiving when it comes to a relatively poor fit for certain feature of a distribution, such as the tail behavior.}

We used \textsc{Stan} to implement the Bayesian inference for the Lake Murray data. {\revise The code can be found in \href{https://github.com/rh8liuqy/flexible\_Gumbel}{\texttt{https://github.com/rh8liuqy/flexible\_Gumbel}}, where the \textsc{JAGS} code for fitting the FG distribution is also provided. The posterior output is given in  Appendix \ref{appendix:convergence_diagnosis}}. The output provided there indicates that our MCMC chain has converged (see the \texttt{Rhat} statistics).

\section{An application in criminology}\label{s:crime}

\begin{table}
	\small
	\caption{\label{tab:crime_modal}Frequentist and Bayesian modal regression models based on the FG distribution fitted to the crime data. Besides parameter estimates (under point.est) and the estimated standard deviations of these parameter estimates (under $\widehat{\mbox{s.d.}}$), 95\% confidence intervals of the parameters from the frequentist method, and 95\% credible intervals from the Bayesian method are also provided (under lower 95 and upper 95).}
	\centering
	\begin{tabular}{lrrrrrrrr}
		\toprule
		\multicolumn{1}{c}{ } & \multicolumn{4}{c}{Frequentist} & \multicolumn{4}{c}{Bayesian} \\
		\cmidrule(l{3pt}r{3pt}){2-5} \cmidrule(l{3pt}r{3pt}){6-9}
		parameter & point.est & $\widehat{\text{s.d.}}$ & lower 95 & upper 95 & point.est & $\widehat{\text{s.d.}}$ & lower 95 & upper 95\\
		\midrule
		$\beta_1$ & -0.166 & 0.072 & -0.306 & -0.026 & -0.162 & 0.079 & -0.312 & -0.003\\
		$\beta_2$ & 0.216 & 0.110 & -0.000 & 0.432 & 0.232 & 0.120 & -0.007 & 0.479\\
		$\beta_3$ & 0.067 & 0.013 & 0.042 & 0.093 & 0.067 & 0.014 & 0.039 & 0.095\\
		$\color{black}\sigma_{1}$ & 1.600 & 0.180 & 1.247 & 1.954 & 1.690 & 0.214 & 1.206 & 2.686\\
		$\color{black}\sigma_{2}$ & 51.882 & 45.034 & -36.384 & 140.148 & 19.3 & 19.300 & 0.187 & 133.047\\
		\bottomrule
	\end{tabular}
\end{table}

\begin{table}
	\small
	\caption{\label{tab:crime_mean}Mean regression model based on the normal distribution fitted to the crime data. Besides parameter estimates (under point.est) and the estimated standard deviations of these parameter estimates (under $\widehat{\mbox{s.d.}}$), 95\% confidence intervals of the parameters are also provided (under lower 95 and upper 95).}
	\centering
	\begin{tabular}{lrrrr}
		\toprule
		parameter & point.est & $\widehat{\text{s.d.}}$ & lower 95 & upper 95\\
		\midrule
		$\beta_1$ & 0.467 & 0.161 & 0.142 & 0.792\\
		$\beta_2$ & 1.140 & 0.224 & 0.689 & 1.591\\
		$\beta_3$ & 0.068 & 0.034 & 0.000 & 0.136\\
		\bottomrule
	\end{tabular}
\end{table}

With the location parameter $\theta$ signified in the FG distribution as the mode, it is straightforward to formulate a modal regression model that explores the relationship between the response variable and predictors. To demonstrate the formulation of a modal regression model based on the FG distribution, we analyzed a data set from \citet{Agresti2021} in the area of criminology. This data set contains the percentage of college education, poverty percentage, metropolitan rate, and murder rate for the 50 states in the United States and the District of Columbia from the year 2003. The poverty percentage is the percentage of the residents with income below the poverty level; the metropolitan rate is defined as the percentage of the population living in the metropolitan area; and the murder rate is the annual number of murders per $100,000$ people in the population. 

We fit the following modal regression model to investigate the association between the murder rate ($Y$) and the aforementioned demographic variables,
$$
Y \mid \boldsymbol{\beta}, \sigma_1, \sigma_2 \sim \operatorname{FG}(\beta_0+\beta_1 \times \text { college }+\beta_2 \times \text { poverty }+\beta_3 \times \text { metropolitan }, \sigma_1, \sigma_2, w),
$$
where $\boldsymbol{\beta} = [\beta_0,\beta_1,\beta_2,\beta_3]^{\top}$ includes all regression coefficients. For the prior elicitation in Bayesian inference, we assume that $\beta_0 ,\dots, \beta_3 \stackrel{\text{i.i.d}}\sim \mathcal{N}(0,10^4)$ and use the same priors for $\sigma_1$, $\sigma_2$ and $w$ as those in  Section \ref{sec:bayes_infer}. As a more conventional regression analysis to compare with our modal regression, we also fit the mean regression model assuming mean-zero normal model error to the data. 

Table \ref{tab:crime_modal} shows the inference results from the modal regression model, and Table \ref{tab:crime_mean} presents the inference results from the mean regression model.  At $5\%$ significance level, both frequentist and Bayesian modal regression analyses confirm that there exists a \textbf{negative} association between the percentage of college education and the murder rate, as well as a positive association between the metropolitan rate and the murder rate. In contrast, according to the inferred mean regression model,  there is a \textbf{positive} association between the percentage of college education and the murder rate. Such claimed positive association is intuitively difficult to justify and contradicts many published results in criminology  \citep{hjalmarsson2012impact,lochner2020education}. 

\begin{figure}[t]
	\centering
	\includegraphics[width = 0.7\linewidth]{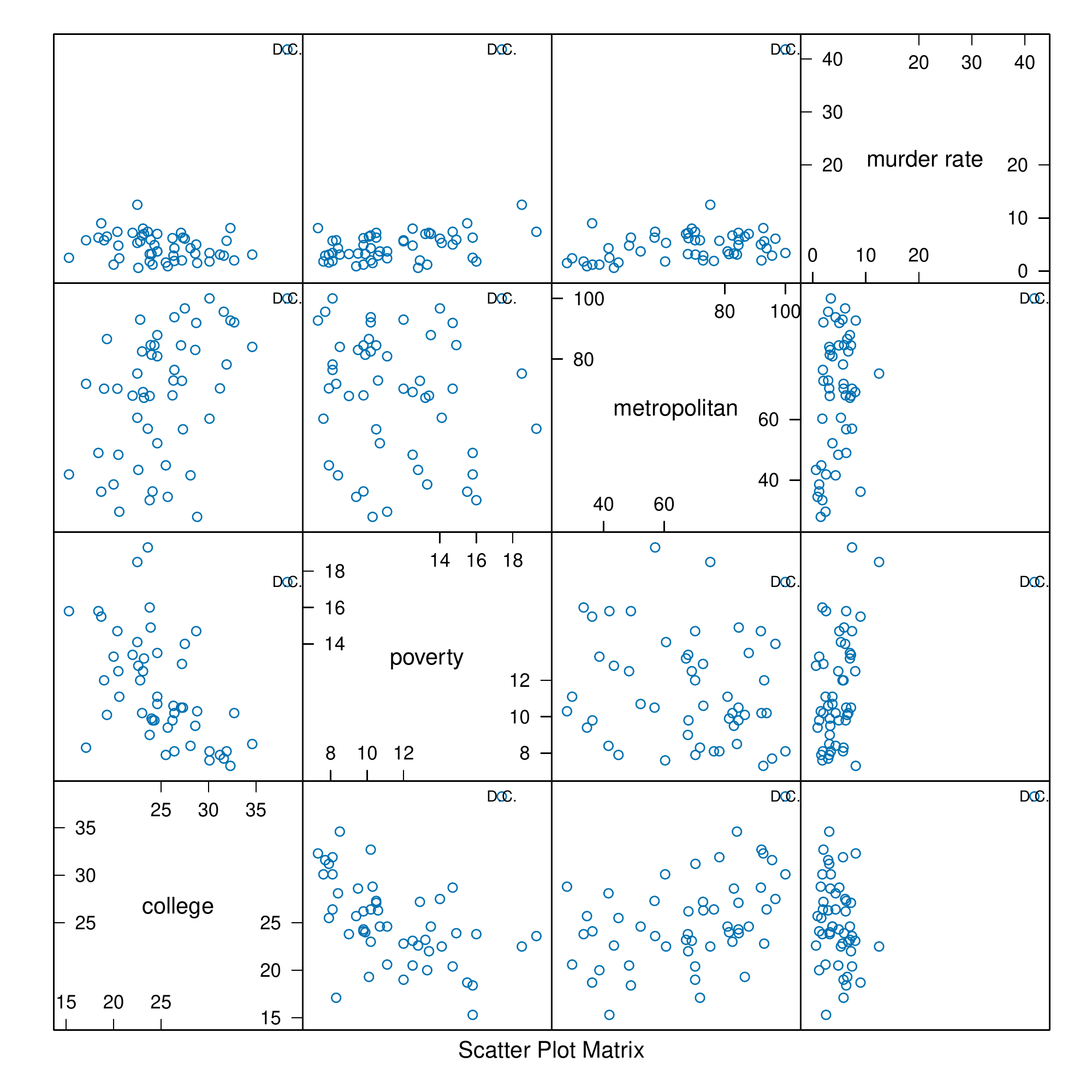}
	\caption{\label{fig:scatter_crime}Scatter plot matrix of the crime data{\revise , where {\color{black} D.C.} stands out as an extreme outlier with the highest murder rate and the highest percentage of college education.}}
\end{figure}

The scatter plot of the data in Figure \ref{fig:scatter_crime} can shed some light on why one reaches such a drastically different conclusion on a covariate effect when mean regression is considered in place of modal regression. As shown in Figure~\ref{fig:scatter_crime}, the exists an obvious outlier, the District of Columbia ({\color{black} D.C.}), in panels of the first row of the scatter plot matrix for instance. {\revise {\color{black} D.C.} not only exhibited the highest murder rate but also the highest percentage of college-educated individuals. These dual characteristics position {\color{black} D.C.}, as an outlier within the dataset.} Mean regression reacts to this one extreme outlier by inflating the covariate effect associated with the percentage of college education in the inferred mean regression function. Thanks to the heavy-tailed feature of the FG distribution, modal regression based on this distribution is robust to outliers, which strives to capture data features suggested by the majority of the data and is not distracted by the extreme outlier when inferring covariate effects in this application. 

{\revise Lastly, to compare their overall goodness of fit for the current data, we computed AIC and BIC following fitting each regression model. Adopting the frequentist and Bayesian methods, the modal regression analysis yields AIC/BIC equal to 239.394/252.917 and 238.710/252.233, respectively. The mean regression analysis leads to AIC and BIC equal to 303.154 and 312.813, respectively. Appendix \ref{appendix:convergence_diagnosis} contains the convergence diagnosis for the Bayesian inferential method applied to this data set, from which we see no concerns about convergence.}

\section{Discussion}
\label{s:discuss}
The mode had been an overlooked location parameter in statistical inference until recently when the statistics community witnessed a revived interest in modal regression among statisticians \citep{chen2018modal, Chacn2020, feng2020statistical, xu2020modal, ullah2021modal, wang2021robust, xiang2022nonparametric}. Historically, statistical inference for the mode has been mostly developed under the nonparametric framework for reasons we point out in Section~\ref{s:intro}. Existing semiparametric methods for modal regression only introduce parametric ingredients in the regression function, i.e., the conditional mode of the response, with the mode-zero error distribution left in a nonparametric form \citep{Yao2013, liu2013robust, zhang2013robust, yang2014robust, zhao2014robust,  krief2017semi, tian2017fitting, li2019linear}. The few recently proposed parametric modal regression models all impose stringent parametric assumptions on the error distribution \citep{bourguignon2020parametric, zhouhuang2020, zhou2022bayesian}. Our proposed flexible Gumbel distribution greatly alleviates concerns contributing to data scientists' reluctance to adopt a parametric framework when drawing inferences for the mode. This new distribution is a heterogeneous mixture in the sense that the two components in the mixture belong to different Gumbel distribution families, which is a feature that shields it from the non-identifiability issue most traditional mixture distributions face, such as the normal mixtures. The proposed distribution is indexed by the mode along with shape and scale parameters, and thus is convenient to use to draw inferences for the mode while remaining flexible. It is also especially suitable for modeling heavy-tailed data, whether the heaviness in tails is due to extremely large or extremely small observations, or both. These are virtues of FG that cannot be achieved by the popular normal mixture and many other existing mixture distributions.  

We develop the numerically efficient and stable ECM algorithm for frequentist inference for the FG distribution, and a reliable Bayesian inference method that can be easily implemented using free software, including \textsc{Stan}, \textsc{JAGS}, and \textsc{BUGS}. Compared with the more widely adopted mean regression framework, the modal regression model based on FG we entertained in Section~\ref{s:crime} shows great potential in revealing meaningful covariate effects potentially masked by extreme outliers. With these advances made in this study, we open up new directions for parametric modal regression and semiparametric modal regression with a fully parametric yet flexible error distribution, and potentially nonparametric ingredients incorporated in the regression function. 


\section*{Disclosure statement}

Computer programs for implementing the FG distribution, related models and data used in this paper are available at \href{https://github.com/rh8liuqy/flexible_Gumbel}{https://github.com/rh8liuqy/flexible\_Gumbel}.

\appendix

\section[\appendixname~\thesection]{\color{black} Derivation of the third central moment of FG in \eqref{eq:m3}}\label{appendix:thired_central_moment}

For any finite mixture distribution, its higher-order central moments can be expressed using the binomial formula (See Equation (1.22) in \citep{fruhwirth-schnatter2006finite}). The FG distribution is a mixture distribution with two components: a right-skewed Gumbel distribution and a left-skewed Gumbel distribution. Let $Y \sim \operatorname{FG} \left(\theta = 0, \sigma_{1}, \sigma_{2}, w\right)$, $Y_{1} \sim \text{right-skewed Gumbel} \left(\theta = 0, \sigma_{1} \right)$, and $Y_{2} \sim \text{left-skewed Gumbel} \left(\theta = 0, \sigma_{2} \right)$. Its $j$-th finite moment can be expressed as
\begin{equation}
	\begin{aligned}
		\mathbb{E}\left\{\left(Y - \mu_y\right)^{j}\right\} &= w \mathbb{E}\left\{\left(Y_{1} - \mu_{1} + \mu_{1} - \mu_y\right)^{j}\right\} + \bar{w} \mathbb{E}\left\{\left(Y_{2} - \mu_{2} + \mu_{2} - \mu_y\right)^{j}\right\} \\
		&= \sum_{k=0}^{j} w \mathbb{E} \left\{ \left(\begin{array}{l}
			j \\
			k
		\end{array}\right) \left(Y_{1} - \mu_{1}\right)^{k} \left( \mu_{1} - \mu_{y} \right)^{j-k}\right\} \\
		& \quad + 
		\bar{w} \mathbb{E} \left\{ \left(\begin{array}{l}
			j \\
			k
		\end{array}\right) \left(Y_{2} - \mu_{2}\right)^{k} \left( \mu_{2} - \mu_{y} \right)^{j-k}\right\},
	\end{aligned}
	\label{eq:appendix_central_moment}
\end{equation}
where $\mu_{y}$, $\mu_{1}$, and $\mu_{2}$ are the expectations of $Y$, $Y_{1}$, and $Y_{2}$, respectively. Applying \eqref{eq:appendix_central_moment} for $j = 3$, one obtains the expression in \eqref{eq:m3}.

\section[\appendixname~\thesection]{\color{black} Proof of identifiability of FG in \eqref{eq:PDFFG}}\label{appendix:identifiability}
In our context of two-component mixture distributions of which the cumulative distribution functions are of the form $wF_1(x)+\bar w F_2(x)$, Theorem 1 in \citet{teicher1961identifiability} states that a mixture distribution is identifiable if and only if there exists $y_1$ in the support of $F_1(y)$ and $y_2$ in the support of $F_2(y)$ such that 
\begin{equation*}
	\left|\begin{array}{cc}
		F_{1}\left(y_{1}\right) & F_{2}\left(y_{1}\right) \\
		F_{1}\left(y_{2}\right) & F_{2}\left(y_{2}\right)
	\end{array}\right|\ne 0,
\end{equation*}
that is, the above determinant does not vanish for some $(y_1, y_2)$. In what follows, we prove that the FG distribution is identifiable by showing the existence of $(y_1, y_2)$ that makes the above determinant non-zero. 
\begin{proof}
	Recall that the cumulative distribution functions of right-skewed and left-skewed Gumbel distributions are given by
	\[
	F_{1}(y) = \exp \left\{-\exp\left(-\frac{y - \theta}{\sigma_{1}}\right)\right\},
	\]
	and
	\[
	F_{2}\left(y\right) = 1 - \exp \left\{-\exp\left(\frac{y - \theta}{\sigma_{2}}\right)\right\},
	\]
	respectively. 
	
	By setting $y_1=\theta$, we have 
	\begin{equation}
		\left|\begin{array}{cc}
			F_{1}\left(y_{1}\right) & F_{2}\left(y_{1}\right) \\
			F_{1}\left(y_{2}\right) & F_{2}\left(y_{2}\right)
		\end{array}
		\right| = \left|\begin{array}{cc}
			e^{-1} & 1-e^{-1} \\
			F_{1}\left(y_{2}\right) & F_{2}\left(y_{2}\right)
		\end{array}\right|=e^{-1}F_2(y_2)-(1-e^{-1})F_1(y_2).
		\label{eq:determinant_appendix}
	\end{equation}
	We next show by contradiction that there exists $y_2\in \mathbb{R}$ such that \eqref{eq:determinant_appendix} is not equal to zero. 
	
	Suppose \eqref{eq:determinant_appendix} is equal to zero for all $y_2\in \mathbb{R}$, that is,
	$$F_2(y_2)=(e-1)F_1(y_2), \text{ for all $y_2\in \mathbb{R}$}.$$
	Taking the limit of both sides of the above equation  as $y_2\to +\infty$ gives 
	$$
	\lim_{y_{2} \rightarrow +\infty} F_{2}\left(y_{2}\right) = (e-1)\times \lim_{y_{2} \rightarrow +\infty} F_{1}\left(y_{2}\right),
	$$
	which is clearly false since $\lim_{y_2\to +\infty} F_2(y_2)=\lim_{y_2\to +\infty} F_1(y_2)=1$. Hence, there exists $y_2 \in \mathbb{R}$ such that \eqref{eq:determinant_appendix} is not equal to zero. Denote by $y_2^*$ such a value, or one of such values if such $y_2$ is not unique.
	
	Now that we have found $(y_1, y_2)=(\theta, y_2^*)$ such that the aforementioned determinant does not vanish, the FG distribution is identifiable by Theorem 1 in \citet{teicher1961identifiability}. 
\end{proof}

\section[\appendixname~\thesection]{\color{black} Convergence diagnosis of MCMC}\label{appendix:convergence_diagnosis}

\begin{Verbatim}[fontsize=\footnotesize]
	## hydrology example - Section 5
	mean se_mean   sd     2.5%      25%      50%      75%    97.5%    n_eff Rhat
	w1         0.61       0 0.14     0.31     0.51     0.63     0.72     0.83 149860.5    1
	theta     -0.42       0 0.70    -1.62    -0.93    -0.48     0.04     1.04 162508.9    1
	sigma1     5.54       0 0.65     4.67     5.11     5.40     5.81     7.25 159083.7    1
	sigma2     5.95       0 1.04     4.58     5.20     5.73     6.49     8.49 175171.4    1
	## criminology example - Section 6
	variable    mean  median      sd     mad      q5      q95  rhat ess_bulk ess_tail
	<chr>      <dbl>   <dbl>   <dbl>   <dbl>   <dbl>    <dbl> <dbl>    <dbl>    <dbl>
	1 alpha     0.496   0.530   2.72    2.68   -4.00     4.89    1.00  102742.   41503.
	2 beta[1]  -0.160  -0.162   0.0785  0.0769 -0.286   -0.0292  1.00   35063.   13152.
	3 beta[2]   0.235   0.232   0.124   0.120   0.0374   0.441   1.00    2074.    1702.
	4 beta[3]   0.0669  0.0669  0.0144  0.0140  0.0434   0.0904  1.00    9603.    9166.
	5 scale1    1.95    1.69    2.36    0.214   1.40     2.64    1.01     588.     189.
	6 scale2   53.6    38.6    74.2    19.3     1.80   133.      1.01     583.     194.
\end{Verbatim}

\bibliography{bibfile}
	
\end{document}